# Electrically tuned super-capacitors

Tazima S. Chowdhury and Haim Grebel*

*Electrical and Computer Engineering and Electronic Imaging Center, New Jersey Institute of Technology, Newark, NJ 07102, USA.*

Abstract:

We show how to increase the capacitance of super-capacitors by modifying the otherwise passive separator layer and turn it into an active diode-like structure. We demonstrate 15% capacitance increase for a diode-like separators without a bias. Additional 15% capacitance increase has been observed for biased separators.

* Corresponding author

E-mail address: grebel@njit.edu

# I. Introduction:

Fast charging and discharging of large amounts of electrical energy make super-capacitors ideal for short-term energy storage [1-5]. In its simplest form, the super-capacitor is an electrolytic capacitor made of an anode and a cathode immersed in an electrolyte. As for an ordinary capacitor, minimizing the charge separation distance and increasing the electrode area increase the overall cell's capacitance. In super-capacitors, charge separation is of the the order of nanometer scale at each electrode's interface (the Helmholtz double layer); making the electrodes porous increases their effective surface area [6-8]. An inherent separating layer between the anode and the cathode electrodes minimizes unintentional electrical discharge (Fig. 1a). We focus on the otherwise insulating but permeable separator and turn it into a permeable and electrically active layer (Fig. 1b-d).

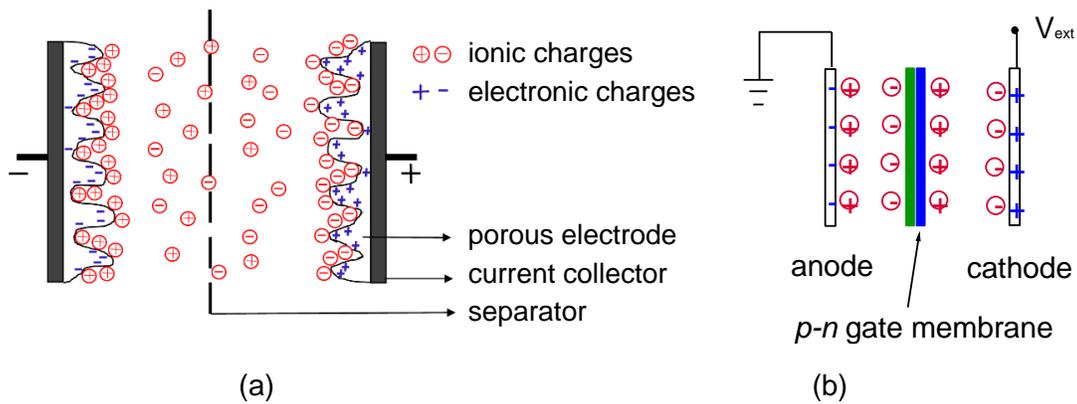

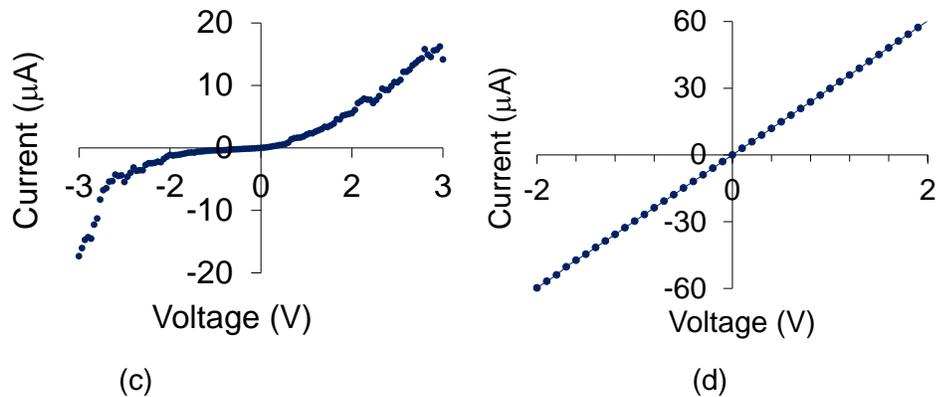

Fig. 1. (a) Schematics of super-capacitors. (b) Diode-like mid-cell gate structure: blue – electronic charges; red – ionic charges; green/blue layers – p-type/n-type layers. (c) Electronic, current-voltage (I-V) curve for a dry p-n gate coating the separator. (d) I-V curve for a single layer (p-type).

The capacitance of a typical parallel plate capacitor is given by: $C=\varepsilon A/d$. Here $\varepsilon$ is the effective permittivity of the cell's electrolyte, $A$ is the area of each electrode and $d$ is the effective charge separation length. Most efforts to date were devoted to optimizing the anode, cathode or the electrolyte. We instead concentrate on the separator layer and modify the otherwise electrically passive layer [9-11] into electrically active diode-like layer [12-14]. Electrically active layers (either diode-like or transistor-like have been also proposed for corrosion protection [15].

Ordinary solid-states electronic diodes are conductive in one direction. The equivalent circuit of such diodes includes capacitive element and it is often employed in large scale integrated circuits [16]. *Our diode-like separators (which are called gates) are permeable to ionic currents* and exhibit both capacitance and resistance to the ion flow. As a result, and as shown below, the overall cell's capacitance is increased. When bias is applied to the gate, the cell's capacitance is further increased, however, at some energy costs.

This paper is organized as follows. We use two techniques to assess our system: Cyclic Voltammetry (CV) and Charge-Discharge (C-D) each used while applying bias to the gate membrane, Vg. Impedance spectroscopy measurements are also provided. Experiments and Methods are provided at the end of this paper.

## II. Results and Discussion:

We used functionalized single wall carbon nanotubes (SWCNT) for the structured *p-n* gate element (Fig. 2) because they exhibit little oxidation [17-21]. As shown in the figure, some of the CNT formed bundles which did not affect their electrical properties under dry conditions (Fig. 1c-d). The two-layer coating, made of numerous junctions was deposited on a hydrophilic polyamide film (TS80) with 0.5-micron pore size. The electrolyte was 1 M NaCl (see also Experiment and Methods section below). The cell's capacitance was determined from 2-electrode experiments [22].

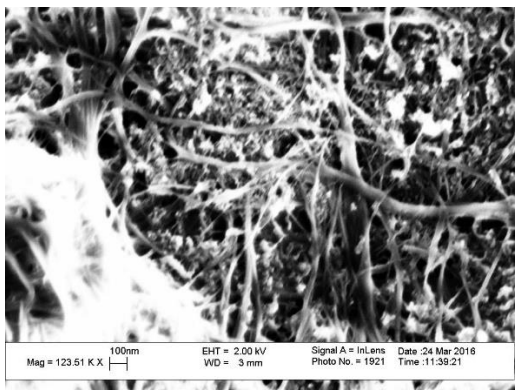

Fig. 2. SEM picture of the diode-like film on top of the separator membrane. The film was made of two layers pressed together: a *p*-type layer and an *n*-type layer each made of functionalized SWCNT.

### II.a. Cell capacitance under no gate bias:

The effect of placing the SWCNT made *p-n* junction(s) on a bare membrane clearly increases the overall cell's capacitance (Fig. 3a-b). CV is sensitive to the capacitance of the entire cell as indicated by the optimal scan rate of 0.2 V/sec. In C-D, one varies the biasing conditions instead of varying the scan rate. Both methods exhibit capacitance increase between 10-18% under no gate bias. C-D experiments consistently exhibit larger capacitance values.

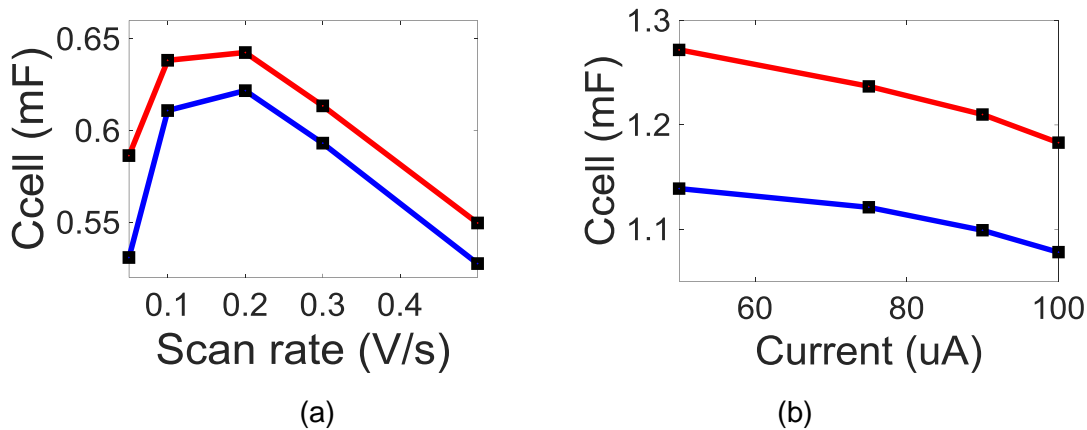

Fig. 3. The effect of a p-n gate membrane on the cell's capacitance using: (a) 2-electrode CV and (b) 2-electrode C-D. Experiments were conducted with no gate bias. The red curve was obtained with p-n gate diode while the blue curve was obtained without it.

**II.b. Cell capacitance under gate bias:**

In Fig. 4a we show results for CV scans while the gate is under voltage bias. The scans were averaged over 4 cycles. Similar curve for C-D is shown in Fig. 4b. Both curves clearly exhibit additional change in the cell's capacitance of ca 15% as a function of gate bias near a gate voltage of Vg~0 V. This additional capacitance comes at some energy cost as we shall see below.

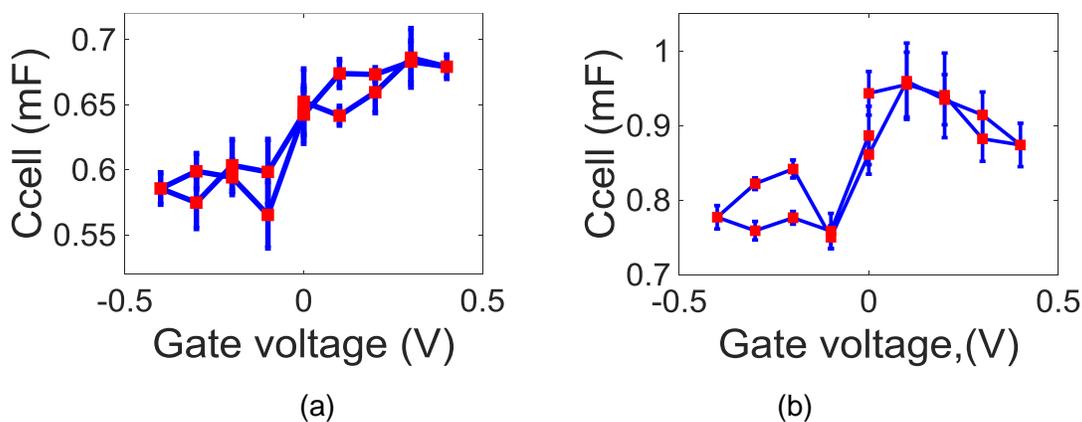

Fig. 4. Capacitance as a function of gate bias, Vg: (a) using 2-electrode CV. (b) Using 2-electrode C-D at current level of 50 micro-Amps. The gate voltage, Vg varied from -0.4 to +0.4 Volts. Vg=0 is the starting point.

Charge separation at the gate is the reason for the observed capacitance change. Under no bias, ionic charge is distributed at the p-n junction(s) as a result of the different electronic chemical potential of the p-type and n-type of the functionalized carbon nanotubes. The polarizing effect at the gate attracts opposite charges to the cell's electrodes and hence, the overall cell's capacitance increases. *Since the CV scans indicate lack of reactions in our experiments*, the Ig-Vg curve shown in Fig. 5 is indicative of the steady state electronic charging-discharging of gate capacitor.

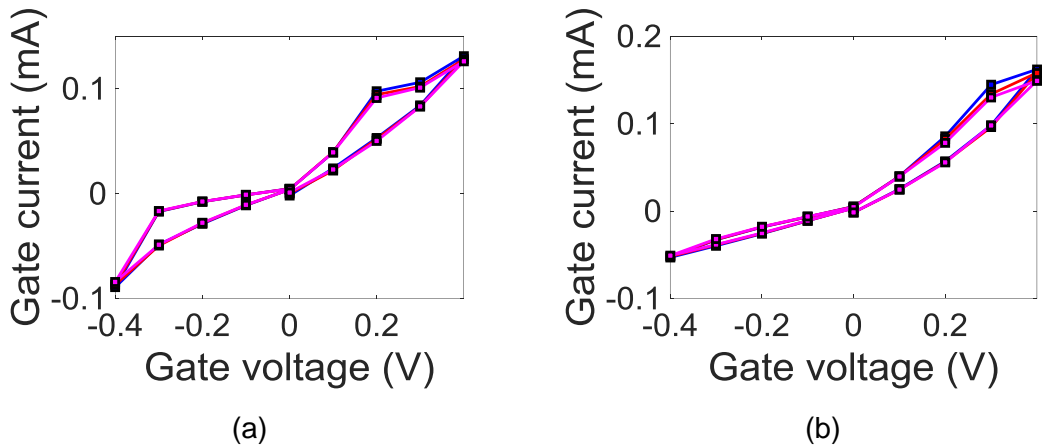

Fig 5. Gate current, Ig as a function of gate voltage, Vg. Data were taken at 30 sec (blue), 40 sec (red) and 50 sec (magenta) while either (a) CV or (b) C-D measurements were running.

**II.c. Energy considerations for a biased gate:**

Biasing the gate creates additional polarizing effect, yet, costs energy. From Fig. 5 and assuming a linear change at Vg~0 V, the largest change in the cell's capacitance under bias occurs for Vg between ±0.1 V, for which $\Delta Ig$~60 µA. The energy invested in charge separation at the biased gate is the area under the Ig-Vg curve, or, $\Delta U_{gate} = \frac{1}{2}\Delta Ig \Delta Vg = 6$ µJ.

Under gate bias, the additional stored energy in our cell may be estimated as follows:

$$U_{cell} = [C_{cell}(V_g)][V_{cell}]^2 \sim \{C_0 + [(\delta C_{cell})/(\delta Vg)](\Delta Vg)\}[V_{cell}]^2 \quad \text{(1a)}$$

$$\Delta U_{cell} = [(\delta C_{cell}/\delta V_g)\Delta V_g](V_{cell})^2 \qquad (1b)$$

$(\delta C_{cell}/\delta V_g) = (0.2m\ F/0.2\ V) = 1 \times 10^{-3}$ F/V; $\Delta V_g = 0.1$ V; $V_{cell}$ is linearly varying between 0 and 0.5 V so, $<V_{cell}> = 0.25$ V. Plugging it into Eq. 1b we get, $\Delta U_{cell} = 6.25\ \mu J \sim \Delta U_{gate}$. Thus, upon biasing the gate, the invested energy is completely utilized in additional energy storage through the increase in the cell's capacitance.

**II.d. Cell model and impedance spectroscopy**

Our cell model is presented in Fig. 6a. The diode is replaced by a capacitor for fitting purposes (Fig. 6b). Nyquist plot and Bode plots are provided in Fig. 6c-d, for a p-n gate membrane at Vg=0 V. Electrochemical impedance spectroscopy was performed from 50 kHz to 100 mHz at 10 mV with a 2 electrode setup. In general, adding a p-n gate structure *reduces* the cell's resistance with respect to its bare counterpart.

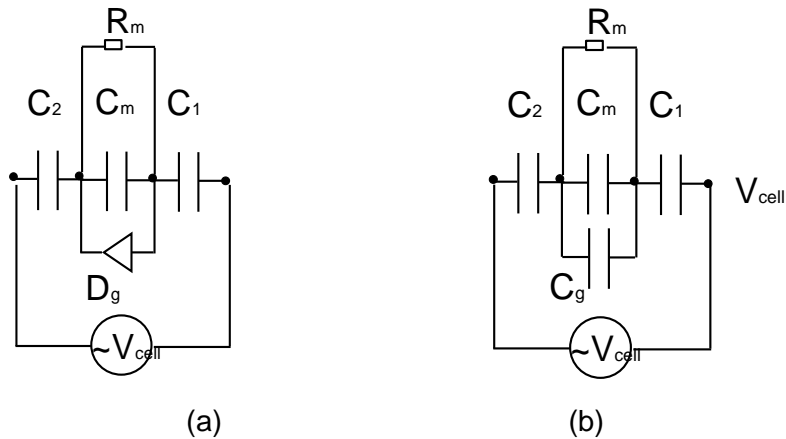

(a)       (b)

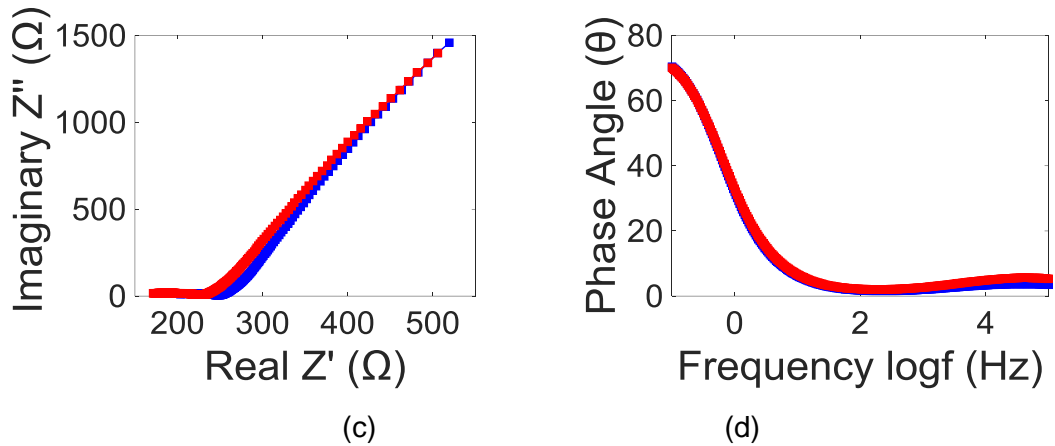

Fig. 6. (a) The equivalent circuit with a permeable diode is replaced by a permeable gate capacitor in (b). $C_1$ and $C_2$ are the double layer interfaces at the cell's electrodes, $C_m$ is the membrane capacitance and $C_g$ is the gate capacitance. (c,d) Electrochemical impedance spectroscopy; bare separator (blue) and a p-n gate membrane as a separator (red) in a 2-electrode cell: (c) Nyquist and (d) Bode plots.

**III. Conclusions:**

In summary, we demonstrated that we can increase and control the capacitance of supercapacitors with active, diode-like separators (the gate). The effect may be separated into two: unbiased gate results in an overall cell's capacitance increase of ca 15%, which is related to ionic charge distribution at the unbiased gate. Further increase in the cell capacitance is achieved upon biasing and therefore polarizing the gate electrode.

**Experiment and Methods:** Aqueous solutions were prepared using deionized water. Single wall carbon nanotubes were obtained from Nano Integris, Canada, with a purity better than 95%. The CNTs were refluxed in dilute nitric acid to remove metal catalyst particles, which are typically used in the growth of CNT. The nitric acid was later washed away and replaced by DI water. The nanotubes were functionalized with polymers: *p*-type and *n*-type tubes were obtained by wrapping the tubes with PVP and PEI, respectively [15]. The *p*-type and *n*-type nanotubes were suspended in DI water using a horn probe sonicator for 8 hours. Each layer was drop-casted on a hydrophilic nano-filtration TS80 filter using vacuum. Copper leads were attached to areas far removed from the electrolyte using silver epoxy. The thickness of each film was estimated as a few microns. Two vertically stripes, one made of n-type and the other made of p-type formed the junction area (see SI section). Two copper electrodes each butt-coupled to either the n-type stripe or the p-type stripe were the electronic leads. They were coated with epoxy to prevent corrosion. As seen from the CV curves (SI section) there is no reaction while scanning the cell under various gate bias.

We did not attempt to increase the surface area of either the anode, or the cathode and so the working and counter electrodes were made of graphite rods with a diameter of 5 mm and 1 M of sodium chloride solution served as an electrolyte. Metrohm PGSTAT potentiostat/galvanostat was used to acquire the cyclic voltammetry curves with a voltage sweep between 0 V to +500 mV at various scan rates with 2 electrode setup. Electrochemical Impedance spectroscopy tests were performed using the FRA32 module of the same system. The frequency ranged from 100 mHz to 50 kHz with a perturbation amplitude of 10 mV. The system also enables charge-discharge measurement.

# Electrically tuned super-capacitors

Tazima S. Chowdhury[a] and Haim Grebel[a,*]

[a]*Electrical and Computer Engineering and Electronic Imaging Center, New Jersey Institute of Technology, Newark, NJ 07102, USA.*

Supplementary Information

2-electrode CV was conducted with 1 M NaCl using bare and p-n structured gate membranes at various scan rates without gate bias. The scan range was between 0 to 0.5 V using graphite rods as working and counter electrodes.

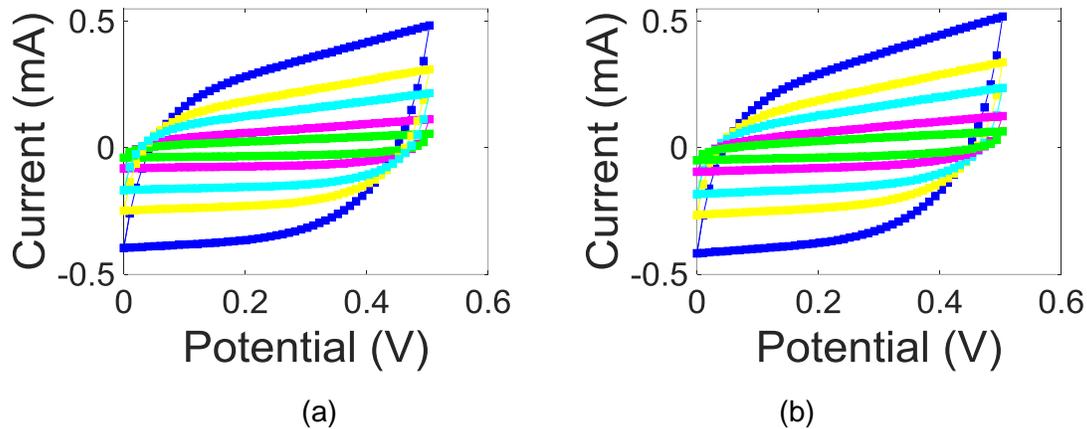

(a)   (b)

Fig. S1. CV curve for various scan rates: 0.05 V/s (green), 0.1 V/s (magenta), 0.2 V/s (light blue), 0.3 V/s (yellow), 0.5 V/s (blue) for both bare (a) and p-n structured separators (b). The slightly larger area in (b) is indicative of the larger cell's capacitor in the case of the p-n structured separator.

The results for a 2-electrode charge discharge (C-D) at various currents are shown in Fig. S2. Voltage range was 0 to +0.5V using graphite rods as working and counter electrodes. Note the slightly longer cycle for the p-n structured separator when compared to the bare separator; this is attributed to the overall cell capacitance increase.

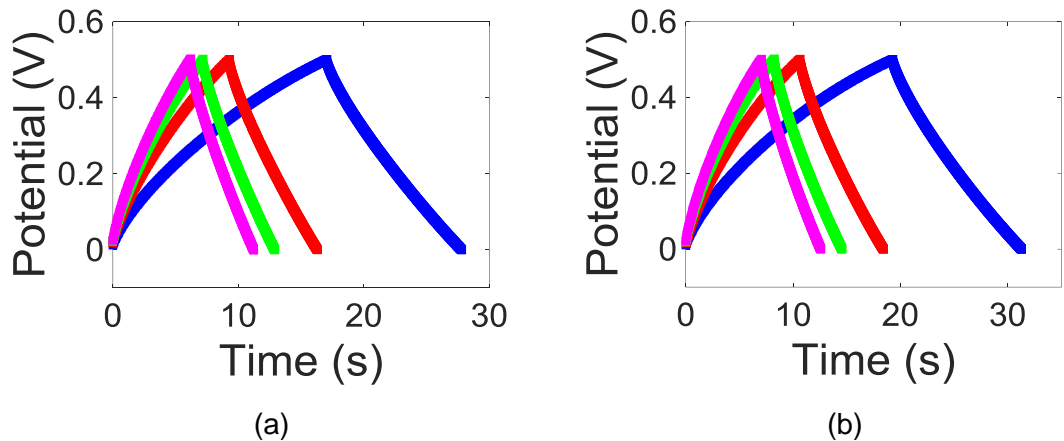

Fig. S2. 2-electrode C-D at various current levels for both bare (a) and p-n structured separators (b): 50 µA (blue); 75 µA (red); 90 µA (green); and 100 µA (magenta). The voltage ranged between 0 and +0.5V. Note the slightly longer cycle for the p-n structured separator due to the overall cell's capacitance increase.

A sample opened after the conclusion of experiments is shown in Fig. S3. Lack of electrode corrosion is noted even after a few weeks of experimentation. Also, the capacitor formed between the two epoxy-coated copper electrode was found to be negligible compared to the gate, formed between the p-type and n-type layers.

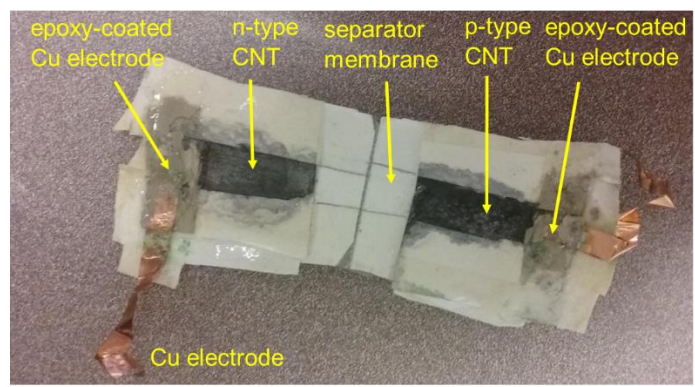

Fig. S3. A sample after the conclusion of all experiments. It did not exhibit corroded leads; in fact, it can be re-assembled and re-used.

In Fig. S4 we show the configuration of our cell. Two graphite electrodes were immersed in 1 M of NaCl. The cell was made of polypropylene. The immovable center plate with a ca 0.3 cm$^2$ hole was made out of PMMA. The diode-like separator was held tight by another PMMA plate with a matching hole to ensure ion flow and a proper contact between the p-type and the n-type coated membranes.

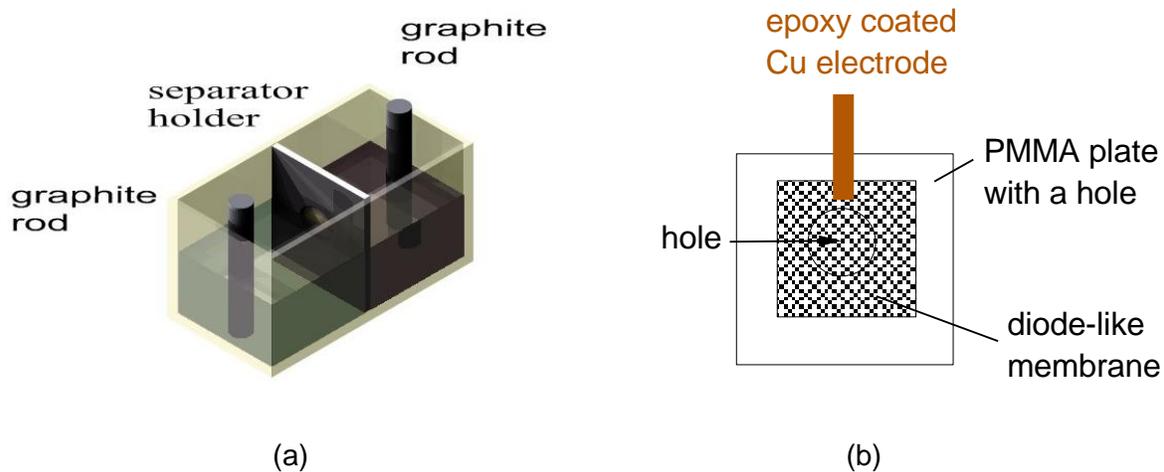

(a)            (b)

Fig. S4. (a) Cell's configuration. (b) The separator holder is composed of two PMMA plates each having a hole. The diode-like separator was held tight between the immovable plate and a removable plate.